\documentclass[12pt, manuscript]{aastex}

\usepackage{amsmath}
\usepackage{color}

\slugcomment{Submitted to ApJ}

\shorttitle{MAGNETIC TOPOLOGY OF CORONAL HOLE LINKAGES}
\shortauthors{TITOV}

\newcommand{\bm}[1]{ \mbox{\boldmath{$#1$}} }
\newcommand{\Rss}{R_{\rm SS}}
\newcommand{\Rs}{R_{\sun}}
\newcommand{\PS}{\Phi_{\sun}}
\newcommand{\PAR}{\Phi_{\rm AR} }
\newcommand{\PDP}{\Phi_{\rm PP} }
\newcommand{\Pds}{\Phi_{\bf -} }
\newcommand{\Pda}{\Phi_{\frown} }
\newcommand{\Pq}{\Phi_{+q}}
\newcommand{\Pqs}{\Phi^{*}_{+q}}
\newcommand{\Pmq}{\Phi_{-q}}
\newcommand{\Pmqs}{\Phi^{*}_{-q}}
\newcommand{\xp}{x^\prime}
\newcommand{\yp}{y^\prime}
\newcommand{\zp}{z^\prime}
\newcommand{\rNn}[4]{$ \left(
   \begin{array}{l} 
     #1 \\ [-13pt]
     #2 \\ [-13pt]
     #3
    \end{array} 
\right)_{\!\!\!\!#4} $}
\newcommand{\rN}[3]{$ \left(
   \begin{array}{l} 
     \scriptstyle{#1} \\ [-15pt]
     \scriptstyle{#2} \\ [-15pt]
     \scriptstyle{#3}
    \end{array} 
\right) $}
\newcommand{\Stoo }{$\rm  BP+N_2+N_1 \:$}
\newcommand{\Stoi  }{$\rm  N_3^{\sun}+N_2+N_1 \:$}
\newcommand{\Stoii }{$\rm  N_3+N_2+N_1 \:$}
\newcommand{\Stoiis}{$\rm  {\hat N}_3+N_2+N_1 \:$}
\newcommand{\Stoiii }{$\rm HFT+N_2^{*}+N_1 \:$}
\newcommand{\Stoiv }{$\rm HFT+N_1 \:$}
\newcommand{\Sti     }{$\rm HFT+N_1 \:$}
\newcommand{\Stii    }{$\rm N_1 \:$}
\DeclareMathOperator{\slog}{slog}
\DeclareMathOperator{\sign}{sign}

\begin{document}

\title{MAGNETIC TOPOLOGY OF CORONAL HOLE LINKAGES} 


\author{V. S. Titov, Z. Miki\'{c}, J.~A. Linker, R.~Lionello}
\affil{Predictive Science Inc., 9990 Mesa Rim Road, Suite 170, San Diego, CA 92121} 
\email{titovv@predsci.com, mikicz@predsci.com, linkerj@predsci.com, lionel@predsci.com}

\and

\author{S.~K. Antiochos} 

\affil{NASA/GSFC, Greenbelt, MD 20771}
\email{spiro.antiochos@nasa.gov}

\begin{abstract}
In recent work, Antiochos and coworkers argued that the boundary between the open and closed field regions on the Sun can be
extremely complex with narrow corridors of open flux connecting seemingly disconnected coronal holes from the main polar holes, and that these corridors may be the sources of the slow solar wind.
We examine, in detail, the topology of such magnetic configurations using an analytical source surface model that allows for analysis of the field with arbitrary resolution.
Our analysis reveals three important new results: First, a coronal
hole boundary can join stably to the separatrix boundary of a parasitic polarity region.
Second, a single parasitic polarity region can produce multiple null points in the corona and, more important, separator lines connecting these points.
It is known that such topologies are extremely favorable for magnetic reconnection, because they allow this process to occur over the entire length of the separators rather than being confined to a small region around the nulls.
Finally, the coronal holes are not {\it connected} by an open-field corridor of finite width, but instead are {\it linked} by a singular line that coincides with the separatrix footprint of the parasitic polarity.
We investigate how the topological features described above evolve in response to the motion of the parasitic polarity region.
The implications of our results for the sources of the slow solar wind and for coronal and heliospheric observations are discussed.

\end{abstract}


%
%
\section{INTRODUCTION
	\label{s:intro}}
A critical issue for understanding the origins and properties of the solar wind is the topology of the magnetic field that connects the corona to the heliosphere, the so-called open field  which we define to be a coronal hole.
In recent work, \citet{Antiochos2007} argued that the boundary between the open and closed field regions on the Sun can be
extremely complex.
In particular, such a boundary has to include narrow corridors of open flux connecting seemingly disconnected coronal holes from the main polar holes, and that these corridors may be the sources of the slow solar wind.
The whole consideration was based on very general theoretical arguments and formulated as the {\it uniqueness} conjecture of coronal holes,
which states that ``coronal holes are unique in that every unipolar region on the photosphere can contain at most one coronal hole" \citep{Antiochos2007}.
On the other hand, observations imply  that coronal holes in some unipolar regions may actually consist of several, apparently disconnected, components (see, e.g., \citet{Kahler2002}).
Similar conclusion also seems to follow from global numerical MHD models of the solar corona \citep{Rusin2010, Linker2010}.

To resolve this discrepancy, first, we construct an analytical model of potential configurations that reproduce the salient features of a numerical magnetohydrodynamic (MHD) model \citep{Linker2010}, in which a moving parasitic polarity region produces an apparent disconnection of the coronal hole.
Then we analyze in detail how the magnetic topology of this field varies in response to the motion of the parasitic polarity.
Our approach relies on the source surface model \citep{Altschuler1969, Schatten1969}, developed here in the exact form for the selected type of configurations.
This allows us to circumvent the common uncertainties of numerical approaches and unambiguously constrain the conditions under which the uniqueness conjecture should be extended to comply with our new findings.

More importantly, our topological analysis of the coronal hole connection and disconnection identifies more accurately the plausible sources of the slow solar wind.
Previously, evidence has been found to suggest that such sources are in a boundary region between coronal holes and active regions  \citep{Ko2006, Harra2008}.
They have also been related to the magnetic reconnection at quasi-separatrix layers (QSLs) with outflows above unipolar regions of the photospheric magnetic field \citep{Baker2009}.
Here we demonstrate that such processes are likely to occur in the course of connection or disconnection of coronal holes by emerging or submerging, respectively, parasitic polarity regions.
In the source surface approximation, this polarity is bordered from disconnecting parts of the hole by a nontrivial combination of genuine separatrix surfaces and QSLs.
A careful analysis of these distinct structural features allows us, first, to understand the topological mechanism of the variation of coronal hole connections and to discover that under certain generic conditions the parasitic polarity has to produce in the corona a so-called separator field line.
It is known (see, e.g., \cite{Lau1990, Priest1996, Longcope2001, Parnell2010}) that this is a likely place for the formation of a strong current layer and magnetic reconnection, which in our case must accommodate the redistribution of magnetic fluxes between closed  and open field structures with significant plasma outflows that can serve as a source of the slow solar wind.
These results complement our other works \citep{Antiochos2010, Linker2010}, where we provide a broader exposition of the relation between the magnetic topology of coronal holes and the slow solar wind.

%
%
%
\section{CONSTRUCTION OF THE FIELD MODEL
	\label{s:mod}}
	
\begin{figure}
\epsscale{0.98}
\plotone{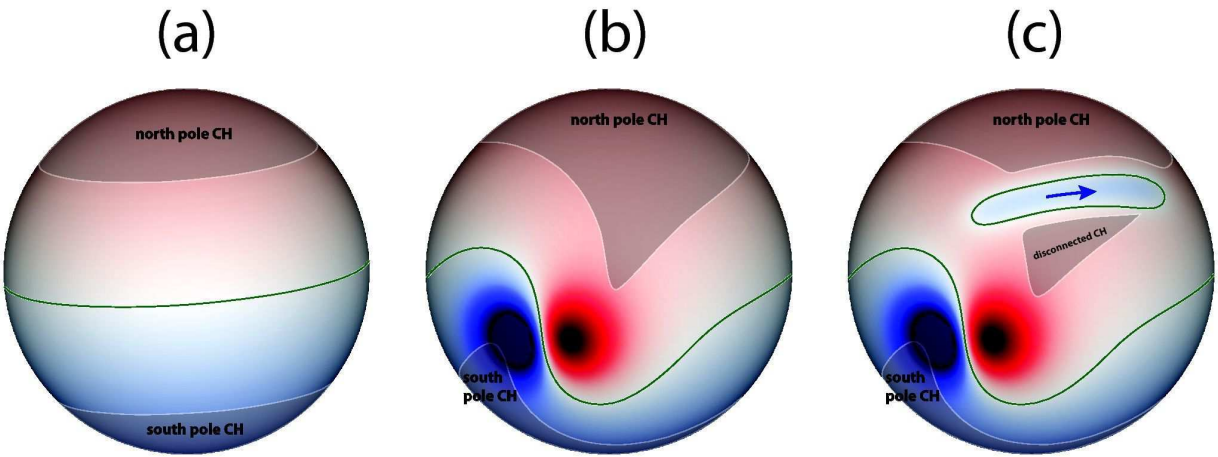}\\
\caption{Three basic steps of constructing the model of magnetic field with a disconnected yet linked (by a separatrix footprint, as shown below) coronal hole (CH):  global large-scale field of the Sun (a) is superimposed with the active region field in order to bulge the pole coronal holes towards the equator (b) and then one of the coronal hole bulges is cut off by adding an elongated negative polarity  in the northern positive hemisphere (c).  The coronal holes are shaded in semi-transparent grey color atop of the respective photospheric red-blue distributions of the radial magnetic field.  
The blue arrow in panel (c) shows the westward direction of the movement of the parasitic polarity that is discussed further in the paper.
	\label{f:f1} }
\end{figure}

Following our numerical MHD model  \citep{Linker2010}, we will construct first the large-scale solar magnetic field that incorporates also a bipole field of an active region.
The incorporated field provides an asymmetry in the shape of polar coronal holes by causing them to bulge towards the flux spots of the same polarity (see Figs. \ref{f:f1}a and \ref{f:f1}b).
Then, in the positive northern hemisphere, we will place at the base of the bulge an elongated negative polarity, which will cut off this bulge into a separate minor hole, as shown in Figure \ref{f:f1}c.
Such a polarity will hereafter be called a {\it  parasitic polarity}.

In our source surface model, we have to construct a potential magnetic field ${\bm B} = - {\bm \nabla} \Phi$ that, first, has no tangential component at the source surface $r=\Rss$, or, equivalently, $\left.\Phi\right|_{r=\Rss}=\mbox{const}$.
Second, the photospheric radial component $\left.B_r\right|_{r=R_\sun}$ must have a certain distribution, which, however, would be sufficient for our purposes to satisfy only qualitatively.
This gives us enough freedom to construct the desirable configuration in a purely analytical form.

Indeed, the linearity of the problem allows us to represent the scalar magnetic potential as
\begin{eqnarray}
  \Phi = \PS + \PAR + \PDP \,,
	\label{Ph}
\end{eqnarray}
where harmonic functions $\PS$, $\PAR$, and $\PDP$ describe, respectively, the global field of the Sun, the active region, and the parasitic polarity.
Figures \ref{f:f1}a,  \ref{f:f1}b, and  \ref{f:f1}c depict the desirable photospheric $B_r$-distributions for $\PS$, $\PS+\PAR$, and total $\Phi$ potentials, respectively.
To satisfy the source surface boundary condition, we will also require that each individual component of the potential must be constant at $r=\Rss$.

Following \citet{Antiochos2007}, we can write the potential of the global field in spherical coordinates $(r,\theta,\phi)$ as
\begin{eqnarray}
   \PS = m \cos\theta \left(\frac{1}{r^2} - \frac{r }{\Rss^3} \right) \, ,
	\label{Ph_sun}
\end{eqnarray}
which is a harmonic function that vanishes at $r=\Rss$.
It is summed from the potential of the dipole $m \hat{\bm z}$ located at $r=0$ and the potential of the uniform field $m/\Rss^3 \hat{\bm z}$.
The unit vector $\hat{\bm z}$ here points in the Cartesian $z$-direction, which passes in our model through the poles of the used spherical system of coordinates.
Hereafter we will assume for simplicity that the $z$-axis is also the rotational axis of the Sun, so that these poles coincide with the north ($\theta=0$) and south ($\theta=\pi$) poles of the Sun, while the east and west directions are to the left and to the right, respectively, from the center of the solar limb.
As will be clear, the fact that in reality the magnetic dipole axis can differ from the rotational axis is not essential for our final conclusions.

To find the active region component, let us apply the method of images \citep{Jackson1962}, so that  the source surface would be equipotential.
This means that the respective potential is decomposed as
\begin{eqnarray}
  \PAR = \Pq + \Pqs + \Pmq + \Pmqs \, ,
	\label{Ph_AR}
\end{eqnarray}
where
\begin{eqnarray}
   \Phi_{\pm q} ({\bm r}) = \frac{\pm q}{\left| {\bm r} - {\bm r}_{\pm q} \right|}
	\label{Ph_q}
\end{eqnarray}
is the potential of fictitious point sources located at ${\bm r}={\bm r}_{\pm q}$ at some depth $d_{\pm q}$ below the photosphere, so that $\left|{\bm r}_{\pm q}\right|=R_{\sun}-d_{\pm q}$.
The potentials $\Phi^{*}_{\pm q}$ of their mirror images can be written in the form
\begin{eqnarray}
   \Phi^{*}_{\pm q}({\bm r}) = -\frac{\Rss}{r} \Phi_{\pm q} \left(  \frac{\Rss^{2}}{r^{2}} {\bm r} \right) \, ,
	\label{Ph_qi}
\end{eqnarray}
which makes it obvious that $\left.\PAR\right|_{r=\Rss}=0$, as required.

This form is actually nothing else than Kelvin's transform [see, e.g., \citep{Axler2001}] applied to the function $\Phi_{\pm q} ({\bm r})$ with the minus sign.
More generally, being applied to {\it any} solution $F({\bm r})$ of Laplace's equation, this transform produces another solution
\begin{eqnarray}
   \tilde{F}({\bm r}) = \frac{R}{r} F\left(  \frac{R^{2}}{r^{2}} {\bm r} \right) \equiv -F^{*}({\bm r}) \, ,
	\label{KT}
\end{eqnarray}
where $R$ stands for the parameter similar to $\Rss$.
This fact is often used for solving electrostatic problems, especially those that involve spherical conductors \citep{Landau1960ecm}, whose analogue in our case is the source surface.
We will apply such a transform twice for constructing the potential $\PDP$.
In principle, the latter could be done even in one step by modeling $\PDP$ with the help of a uniformly charged circular arc and its mirror image.
Unfortunately, the potential of such an arc is complexly expressed in terms of elliptic functions, which provides an essential impediment to topological analysis of the resulting field, and, specifically, to determining magnetic null points and their properties.

To remain within the class of elementary functions, let us use the powerful machinery of Kelvin's transform.
First of all, note that the transform defined by equation (\ref{KT}) can be viewed as a composition of two simpler transforms, one of which is just a nonuniform radial scaling by the factor $R/r$, while the other is a pull-back mapping of the original potential via the sphere inversion
\begin{eqnarray}
   {\bm r} \rightarrow  \frac{R^{2}}{r^{2}}  {\bm r} .
	\label{SI}
\end{eqnarray}
Recall now that the inversion generally maps lines into circles.
This means that if the original potential is singular at some line segment, the transformed potential will be singular at the arc to which the line segment is mapped by equation (\ref{SI}).
Thus, if we start from the potential of a stick with a uniform line distribution of dipoles and subject this potential to Kelvin's transform, we will obtain the potential of an arc with a certain distribution of dipoles and charges along it.
It turns out that the resulting field may perfectly model the required parasitic polarity.

One can easily check that if we place the original stick at the distance $R=2(\Rs-d_{a})$ tangentially to the inversion sphere of radius $R$, we will get the arc of radius $\Rs - d_{a}$.
Bearing this in mind, let us find first the potential $\Pds({\bm r})$ of the stick of length $2l$ by simply integrating the potential $\mu z/\left| {\bm r} - {\bm r}_{0} \right|^3$  of a dipole at ${\bm r}_{0} = (x_0,0,R)$ over $x_{0}$ from $-l$ to $l$; this yields 
\begin{eqnarray}
   \Pds({\bm r}) = 
                                   \frac{\mu z}{|{\bm r}-{\bm r}_{+l}| \left( x-l + |{\bm r}-{\bm r}_{+l}| \right)}
                                  -\frac{\mu z}{|{\bm r}-{\bm r}_{-l}| \left( x+l + |{\bm r}-{\bm r}_{-l}| \right)} \, ,
	\label{Ph_st}
\end{eqnarray}
where ${\bm r}_{\pm l}=(\pm l,0,R)$.
Applying now the transform (\ref{KT}) to this potential at $R=2(\Rs-d_{a})$, we will get the potential $\tilde{\Phi}_{\bf -}({\bm r})$ of the arc of radius $R/2=\Rs - d_{a}$ that is located in the plane $y=0$ at $z=R$, so that the center of the arc is at  $z=R/2$ rather than at the origin of the system of coordinates, as needed.
However, we can easily bring it to the proper place by simply shifting the system of coordinates on the distance $\Rs-d_{a}$ in the $z$-direction.
Then, combining this shift with a suitable rotation of the system of coordinates, we get
\begin{eqnarray}
  && \Pda(x,y,z)  \equiv  \tilde{\Phi}_{\bf -}(\xp,\yp,\zp) \, ,  \\
  && \xp = x \sin\phi_{a} - y \cos\phi_{a} \, ,  \\
  && \yp = x \cos\theta_{a} \cos\phi_{a} + y \cos\theta_{a} \sin\phi_{a} - (z+\Rs-d_{a})\sin\theta_{a} \, ,  \\
  && \zp = x \sin\theta_{a} \cos\phi_{a} + y \sin\theta_{a} \sin\phi_{a} + (z+\Rs-d_{a})\cos\theta_{a} \, ,
	\label{Ph_da}
\end{eqnarray}
which is the required potential of the arc such that $r=\Rs-d_{a}$, $\theta=\theta_{a}$, and $\phi_{a}-\alpha \le \phi \le \phi_{a}+\alpha$ , where $\alpha=\arctan[l/2(\Rs-d_{a})]$.

The analysis of the obtained solution shows that the arc contains not only dipoles but also charges, which are both non-uniformly distributed along the arc.
Although the form of these distributions is not essential for further consideration, we will still need to know the total charge of the arc for adjusting our model to have its total photospheric flux balanced.
We can find the arc charge by calculating the leading term in the asymptotic expansion of $\tilde{\Phi}_{\bf -}({\bm r})$ by large $r$: this term is proportional to  $r^{-1}$ with the coefficient
\begin{eqnarray}
  q_{l} \equiv -\frac{2\mu l}{\left[ l^2+4\left( \Rs - d_{a} \right)^2 \right]^{1/2}} \, ,
	\label{ql}
\end{eqnarray}
which is exactly the required total charge of the arc.

Now we complete the construction of our model by writing the potential of the parasitic polarity as
\begin{eqnarray}
    \PDP({\bm r}) = \Pda({\bm r}) + \Phi^{*}_{\frown}({\bm r}) - \frac{q_{l}}{r} +  \frac{q_{l}}{\Rss} \, ,
	\label{Ph_DP}
\end{eqnarray}
where the potential $\Pda^{*}({\bm r})$ of the arc image is determined by equation (\ref{KT}) with $F$ changed to $\Pda$ and $R$ to $\Rss$.
We have also added here the potential of the fictitious charge $-q_{l}$ placed at the center of the Sun to compensate the indicated charge of the arc and make the total photospheric flux balanced.
The constant parameter $q_{l}/\Rss$ is added to this expression only for esthetics: it makes $\PDP$ to be equal to zero at $r=\Rss$ rather than simply constant.

Thus, equations (\ref{Ph})--(\ref{Ph_qi}) and (\ref{Ph_st})--(\ref{Ph_DP}) fully determine a source surface configuration with a desirable magnetic flux distribution at the photosphere.
The strengths of the sources generating global, active region and parasitic polarity fields are controlled by the parameters $m$, $q$, and $\mu$, respectively.
The widths of the active region spots and parasitic polarity are regulated by the depths $d_{\pm q}$ and $d_{a}$ of the charges and arc, respectively,  while the length of the parasitic polarity is roughly proportional to $2l$.
Finally, the spherical coordinates of the charges $(\theta_{\pm q},\phi_{\pm q})$ and the center of the arc $(\theta_{a},\phi_{a})$ control the locations of corresponding polarities on the solar globe.

%
%
%
\section{BASIC TOPOLOGICAL STATES
	\label{s:topsts}}

Taking the gradients of the potentials described in the previous section, we have calculated the modeled magnetic field.
Then varying the model parameter $\phi_{a}$, with other parameters
being fixed, we have determined a sequence of configurations that
represents the variation of linkages between two coronal holes
in the result of moving parasitic polarity.
We start from the reference state where the holes are {\it disconnected} (Fig. \ref{f:f1}c) but {\it linked} at the photosphere by a singular line (see Figs. \ref{f:f2} and \ref{f:f3}) representing the footprint of a separatrix surface.
Then we gradually convert this singular line into an open-field corridor containing finite flux by moving the parasitic polarity westward (see Fig.  \ref{f:f1}c), so that the holes eventually become {\it connected} by this corridor.
The singular linkage between coronal holes in the reference state is a key new result of this paper, whose implications are discussed in more detail in Section \ref{s:un} below.

Table \ref{t:1} summarizes the properties of the basic topological states through which the configuration passes in this process, starting from the indicated reference state.
The states are named with the lists of the features that constitute the structural skeleton of the respective configuration in the vicinity of the parasitic polarity.
First of all, these are magnetic null points $\rm N_1$,  $\rm N_2$,
 and  $\rm N_3$, whose number changes during the conversion of linkage to connection of the coronal hole via merging of two nulls $\rm N_2$ and  $\rm N_3$ into a degenerate null $\rm N_2^*$ and its subsequent disappearance, or its bifurcation back into $\rm N_2$ and $\rm N_3$ in the reverse process.
At certain values of $\phi_{a}$, the null point $\rm N_3$ may transform into a so-called ``bald patch" (BP), which is a segment of the polarity inversion line (PIL), where a set of coronal magnetic field lines touches the photosphere \citep{Seehafer1986a, Titov1993}.
Our configuration may also have hyperbolic flux tubes (HFTs) \citep{Titov2002}, which are combinations of two intersecting QSLs introduced by \citet{Priest1995} and \citet{Demoulin1996}.
Thus, the acronyms ${\rm N}_{i}$ ($i=1,2,3$), BP, and HFT enter into the names of eight basic states shown in the first column of  Table \ref{t:1}.

The second column of the table presents the numbers of Figures, in which the respective magnetic field structures are depicted.
The third column provides the corresponding values of $\phi_{a}$, and the next three columns give the spherical $(r,\theta,\phi)$ coordinates of the nulls.
These coordinates, as well as the model parameters (see the caption to Table \ref{t:1}),  are rounded to five significant digits.
The lengths are given in units of $R_{\sun}$, while the units of other dimensional parameters are chosen, assuming that the calculated magnetic field is measured in Gauss.
Finally, the last column in the table indicates two properties of a given state: first, whether the northern coronal hole is connected (C) or only linked (L), and, second, whether this state is topologically stable (S) or unstable (U).

%
%
\subsection{Reference State  \Stoo with a Disconnected-Linked Coronal Hole}

The reference state \Stoo with a disconnected, but linked, bulge of the northern coronal hole is characterized by the presence of one BP and two nulls $\rm N_2$ and $\rm N_1$ (Fig. \ref{f:f2}).
Both BP and $\rm N_2$ belong to the fan separatrix surface that emanates from the null $\rm N_1$.
Hereafter we use the terms ``fan surface" and ``spine line" as they were defined by \citet{Priest1996} through the eigenvectors of the matrix of magnetic field gradients at the null points.
The fan surface is woven of the field lines that start at the null point in the plane spanned on the eigenvectors, whose eigenvalues have the same sign, while spine line emanates from the null point along the remaining third eigenvector.
The fan surface associated with the null $\rm N_1$ has a dome-like shape covering the parasitic polarity from above (Figs. \ref{f:f2} and \ref{f:f3}).
This is a typical structure for an isolated polarity region immersed in a dominating flux region of opposite sign.
Such a surface is hereafter called  for brevity the {\it separatrix dome} or simply {\it dome}. 

The appearance of the BP at the eastern side of the parasitic polarity can be understood if one takes into account the prevailing contribution of the active-region flux spots into the local field of the BP.
It overrides the contributions of the global background field and parasitic polarity, thereby turning the vectors of the resulting local field outward from the polarity, which in turn implies the existence of the BP at the respective part of the inversion polarity line \citep{Titov1993}.
Note also that the field orientation at the BP is opposite to the arrow orientation of the field line that goes out from the null $\rm N_1$ towards the BP, which means that the field direction becomes reversed on this line.
Since it is almost a straight line parallel to the parasitic polarity, such a reversal may occur only at a null point.
This provides an explanation of the presence of the null point $\rm N_2$ in the configuration under study.

The local analysis of the eigenvectors at the null $\rm N_2$ shows that its fan separatrix surface is oriented vertically and along the parasitic polarity.
We will call such a surface the {\it separatrix curtain}.
It intersects with the separatrix dome along the field line (thick scarlet line in Figs.  \ref{f:f2} and  \ref{f:f3}) that comprises two curve segments smoothly joined to each other.
The first curve segment connects the nulls N$_1$ and N$_2$ and represents the so-called {\it separator field line} \citep{Baum1980}.
The second curve segment connects the null N$_2$ and the BP by touching the latter at one of its points, from where it smoothly continues as a BP separatrix field line (thick magenta line in Figs.  \ref{f:f2} and  \ref{f:f3}) whose second footpoint locates at the big negative flux spot of the active region.
Strictly speaking, only the western part of the separatrix dome emanates from the null N$_1$, because all field lines of the dome (blue lines in Fig. \ref{f:f3}) that go eastward along the separator eventually deflect from the null N$_2$ and go down to the photosphere along the spine line of the null N$_2$.
At this spine line, the western part of the dome smoothly joins the eastern part, which is entirely formed by the field lines touching the BP.
Therefore, the second curve segment is not a classical ``null-null" field line but rather its an interesting hybrid with the BP separator \citep{Bungey1996}.
In principle, the magnetic reconnection along the indicated two curve segments may behave differently, but since they smoothly join each other, we will consider them as a single entity called for brevity the separator field line.

To make a comprehensive analysis of the magnetic structure, we have computed for our field model the so-called squashing degree or factor $Q$, which was proposed first for describing closed magnetic configurations in Cartesian geometry by \citet{Titov1999a} and \citet{Titov2002}.
The $Q$ factor characterizes the divergence of magnetic field lines, so that its high values identify QSLs and separatrix surfaces in a given configuration.
Here we used a generalized definition of the $Q$ factor \citep{Titov2007a} that is applicable to both closed and open magnetic fields defined in spherical coordinates.
At genuine separatrix surfaces, the $Q$ factor formally tends to infinity, but numerically remains finite, such that its high values grow with decreasing the size of the numerical grid used for calculating $Q$.
In all our plots of the $Q$ distribution at the photosphere and source surface, we have saturated its scale at $Q=10^3$, which is sufficient for our purposes.

As expected, the photospheric high-$Q$ lines trace all the boundaries of coronal holes as well as the footprint of the separatrix dome (see Fig. \ref{f:f2}a, \ref{f:f2}c).
At the source surface, the high-$Q$ lines trace the neutral line and the footprint of the separatrix curtain, which are represented in Figure \ref{f:f2}b by lower and upper arcs, respectively.
These arcs encircle an eye-like area that corresponds to the disconnected bulge of the northern coronal hole.
The upper and lower areas correspond here to the remaining part of the northern coronal hole and the southern coronal hole, respectively.
Figure \ref{f:f2}c illustrates the separatrix curtain that borders the disconnected parts of the northern coronal hole.
The $Q$ distribution shows also that the separatrix curtain and dome are surrounded like a halo by QSLs.
This effect results from a rapid divergence of the field lines in the neighborhood of the null points associated with these separatrix surfaces.

Let us discuss now why the northern coronal hole is not connected at the photospheric level.
Could it be instead still connected via a sort of singular corridor formed by the eastern or western part of the footprints of the separatrix dome?
For the eastern part, we can unequivocally answer ``no", because all the field lines starting at this part are closed (see purple lines reaching the BP in Fig. \ref{f:f3}).
The only exception here is the separator that connects to the null N$_2$ and then to the source surface through the field lines of the separtrix curtain.
Thus, only the single point at which the separator touches the BP can be qualified in the eastern part of the dome footprint as an open field ``region", while the rest of this part belongs to a closed field region.

For the western part, the answer is more subtle, because all the field lines starting at this part enter first the null N$_1$ (see blue lines in Fig. \ref{f:f3}) and then continue their path ``at will", either along the closed spine line or along the separator connecting to the null N$_2$, from where the field lines finally run away to the source surface.
In other words, due to this ambiguity and ``multi-stepness" of the connectivity, we cannot consider the field lines starting at the western part of the dome footprint as definitely open or closed.
Following the field lines that start at the western part of the dome footprint, however, one can always reach via the separator the null point N$_2$ and so the open field lines.
Consequently, there does exist a topological connection between the
coronal holes, but it is not via any finite amount of open flux and,
therefore, it seems inappropriate to label them as ``connected''. 
This motivates us to call such states with a formally disconnected coronal hole as ``linked", which is described in more detail in section \ref{s:un}.

%
%
\subsection{Converting Coronal Hole Linkage into Connection and Back}

Moving down in Table \ref{t:1} from the top to bottom row and looking at the corresponding Figures \ref{f:f2}--\ref{f:f8}, one can follow the variation of the magnetic topology in our model and, in particular, the merging of the disconnected parts of the hole back into a unique hole in response to the westward displacement of the parasitic polarity.
In this process, first, the BP turns gradually into a null point N$^{\sun}_3$ located exactly at the PIL (Fig. \ref{f:f4}),
so that the original BP separatrix field line that smoothly extends the separator in the reference state turns now into the spine line of the null N$^{\sun}_3$.
Locally, we deal here with the case where the BP plays the role of a precursor for an emerging null point \citep{Bungey1996}.
This state \Stoi is topologically unstable, because a small perturbation of the field configuration will cause either the transformation of the null N$^{\sun}_3$ back into a BP or its emergence into the chromosphere and corona.
The latter occurs, in particular, when the parasitic polarity continues moving westward, which leads to a sequence of states \Stoii with three coronal null points, all distributed along the separator (Fig. \ref{f:f5}).
The fan planes associated with the nulls N$_3$ and N$_1$  are both oriented horizontally and tangentally to the separatrix dome.

For the state \Stoi  (Fig. \ref{f:f4}) and its neighboring states of the type \Stoii  (Fig. \ref{f:f5}), the spine lines coming out of the nulls N$^{\sun}_3$ and N$_3$ are closed.
However, with moving the parasitic polarity further westward, the spine apex rises more and more until it touches the source surface.
This occurs precisely at the eastern cusp of the eye-like contour traced by the high-$Q$ lines at the source surface (seen in Fig. \ref{f:f5}b).
Starting from this moment, such a spine line becomes open together with all the field lines entering the null $\rm \hat{N}_3$ from the eastern part of the dome footprint.
Thus, right at this moment, this part of the footprint can be considered as a singular corridor that links the disconnected coronal hole, which is indicated in Table \ref{t:1} as a singularly connected ($\rm \hat{C}$) state \Stoiis.
Further displacement of the parasitic polarity to the west extends such a corridor to a finite width.

A precise calculation of the parameters of the state \Stoiis is not a simple problem, since it implies the determination of the respective $\phi_{a}$ from the nonlocal condition requiring that the N$_3$-associated spine must hit the neutral line at the source surface.
We did not solve this nontrivial problem, because for our purposes, it is sufficient to realize the mere existence of the state \Stoiis.
The latter, however, undoubtedly follows from the continuity of the model by parameter $\phi_{a}$ and the presence of the states with closed and open spine lines emanating from the nulls N$_3$.
It will be clear from the discussion below and Table \ref{t:1} that the corresponding value of $\phi_{a}$ lies in the interval $[3.76,\: 3.79]$. 

Starting from the state \Stoiis, further westward movement of the parasitic polarity causes the disconnected parts of the coronal hole to merge with each other through a widening corridor and form a unique coronal hole.
All three of the nulls move together with the parasitic polarity in the same direction but with different velocities such that the null N$_3$ moves faster than the two others and eventually catches up and coalesces with the null N$_2$.
At this moment, the configuration reaches a new type of topological states (Fig. \ref{f:f6}) denoted as \Stoiii.
It is characterized by the presence of an HFT and a degenerate null point N$^*_2$ such that one of its eigenvalues identically vanishes.
The photospheric footprint of the HFT in this state is an extremely narrow but finite-width corridor that connects the initially disconnected coronal hole.
Its source-surface footprint is shown in Figures \ref{f:f6}b,  \ref{f:f6}c, and  \ref{f:f6}d by dashed lines going along the eastern part of the upper high-$Q$ arc of the eye-like contour.

The state \Stoiii is topologically unstable, since a small displacement of the parasitic polarity back to the east leads to the bifurcation of the null N$^*_2$ into a pair of nulls N$_2$ and N$_3$, while its displacement further to the west causes a full disappearance of N$^*_2$.
However, disappearing as a topological feature, the null N$^*_2$ ``reincarnates" as a geometrical feature into a local minimum point  of $|\bm B|$.
Following to \citet{Priest1996b}, we could call the inverse process as the {\it saddle-node-Hopf bifurcation} of a magnetic minimum.
It is worth also to mention that after converting N$^*_2$ into a magnetic minimum the separator field line disappears becoming just one of the ordinary separatrix field lines of the dome.

With moving the parasitic polarity further to the west, the configuration passes through a sequence of states \Stoiv that exist for a wide range of $\phi_{a}$.
The characteristic feature of these states is the presence of an HFT associated with the indicated magnetic minimum, in whose neighborhood the field lines experience a big divergence.
Taken as a whole, the HFT consists of two QSLs, one of which skirts the separatrix dome, while the other hangs above the dome in place of the previous separatrix curtain (Fig. \ref{f:f7}).
They join one another into the HFT and form a T-type junction at the top of the dome along the field line, which we will call the {\it quasi-separator} by analogy with similar field lines in quadrupole configuaritons \citep{Titov2002}.
As clearly seen in Figure \ref{f:f7}a, the vertical QSL actually prolongs below the dome and creates inside its oval footprint a well distinguished high-$Q$ line, which can be viewed as a characteristic photospheric signature indicating the presence of the quasi-separator at the dome.
By analogy with the separator field line, we think that the quasi-separator must be a preferred site for the formation of a thin current layer and reconnection during the MHD evolution of the configuration.
In other words, in spite of essential differences in topology of the magnetic field, the configurations with quasi-separatrix and true separatrix curtains must response similarly to MHD perturbations.

Right after the disappearance of  the degenerate null N$^*_2$, the source-surface footprint of the HFT is very similar to the respective footprint of the separatrix curtain at previous states (cf. panels (b) in Figures \ref{f:f2}--\ref{f:f7}).
Yet with moving the parasitic polarity further to the west and widening the corridor between initially disconnected parts of the hole, this footprint shrinks in the western direction along with the associated HFT (see Fig. \ref{f:f8}).
The magnetic minimum becomes in this process  more and more shallow and subsequently disappears.
This is a particular manifestation of a more general relationship between magnetic minima and QSLs pointed out recently by \citet{Titov2009}.
Thus, figuratively speaking, the restoration of the photospheric connection in the coronal hole is accommodated in the corona, first, by the transformation of the separatrix curtain into a quasi-separatrix one and, then, by its gradual opening.

The process of the disconnection of the northern coronal hole is recovered in our model by simply following the states presented in Table \ref{t:1} in reverse order, starting from the state \Stii at the bottom of the table.

%
%
\subsection{Extension of the Uniqueness Conjecture \label{s:un}}

\citet{Antiochos2007} recently derived a uniqueness conjecture stating that the parts of coronal holes within unipolar photospheric regions remain connected at all times.
It was noted though that, in some cases, coronal holes are connected via extremely narrow corridors.
Our examples confirm this important conclusion and motivate its further extension by demonstrating that, in fact, such corridors can even shrink to singular lines of zero width.
In these cases, the different parts of coronal holes formally lose
their connection in the photosphere, since such singular corridors
have open magnetic flux of measure zero.
Therefore, the uniqueness conjecture needs to be refined to describe this new type of configuration.  

We do this by extending the definitions of coronal hole connectedness as follows.
We call two parts of a coronal hole {\it disconnected} in the
photosphere if there is no corridor with a finite open magnetic flux
that connects these parts.
Thus, two coronal holes, such as those depicted in Figures \ref{f:f2}--\ref{f:f5},
that are joined by a zero-width footprint of a separatrix dome, are considered
to be disconnected, since this footprint has zero magnetic
flux.
Certainly, such a coronal hole configuration would appear to be
observationally disconnected.
To distinguish these special states from coronal holes that are connected
via a finite-width corridor in the photosphere, we define the term
{\it linked} to describe this kind of singular ``connection'' by a line
with zero magnetic flux.
Therefore, in the sequence of states that we have considered, the two
pieces of a coronal hole can first be connected, and then become
disconnected, but remain linked.
It is apparent now that the idea behind the original statement of the uniqueness conjecture \citep{Antiochos2007} was sound, but its justification was not entirely correct.
We have addressed this by defining the connectedness of coronal holes in a broader sense, as stated above.
To be precise, we restate the uniqueness conjecture to say that coronal holes in unipolar photospheric regions are always either {\it connected} or {\it linked}, or {\it both}.

It is important to understand exactly why the arguments for uniqueness
presented by \citet{Antiochos2007} fail for the magnetic topologies
studied in this paper. Note that our photospheric flux distribution is
quite simple, consisting of a global dipole and a single parasitic
polarity region, essentially identical to that in \citet{Antiochos2007}.
The key difference in our study, however, is that the topology
associated with the parasitic polarity, for example, the field of
Figures \ref{f:f2} and \ref{f:f3}, is more complex than that considered by Antiochos et al.
Those authors assumed the simplest possible, and most common, topology
consisting of a dome-separatrix surface with a single null and two
spine lines. For this topology, the position of the null completely
defines whether the parasitic polarity is inside the closed field or
the open field region. The situation where the null point is exactly
on the open-closed boundary surface, so that the parasitic polarity
separatrix curve on the photosphere coincides over part of its length
with the coronal hole boundary, is a singular case that is
structurally unstable. Any perturbation at the photosphere will move
the null and break the degeneracy between the separatrix curve and the
coronal hole boundary. 
Consequently, \citet{Antiochos2007} concluded
that ``a nested polarity region must be surrounded by either all open
or all closed field''.

In our case, however, the parasitic polarity topology has multiple
null points, in fact, the number can change due to saddle-node-Hopf
bifurcations, along with a separator line connecting the nulls. This
allows one of the nulls to remain stably on the open-closed boundary
surface and the photospheric separatrix curve to remain degenerate
with the coronal hole boundary.  We find, therefore, that the
parasitic polarity region is not surrounded by all open or all closed
field, but is actually bounded by both.
Note that in Fig. \ref{f:f3}, the separatrix curve on the photosphere, orange ellipse surrounding the green PIL, coincides with the
coronal hole boundary at the two points P$_1$ and P$_2$.
As a result, the connection between the upper and lower
coronal holes is no longer via a finite-flux open corridor, but via
this parasitic polarity separatrix curve. Furthermore, this singular
linkage is structurally stable to finite changes at the photosphere.

It is worth remembering that even though the configurations with finite-width and zero-width corridors are very distinct topologically, their impact on the physical processes in the corona may not be so different.
As can be seen from our examples,  in both these types of configurations, the bulk of the field lines with rapidly varying connectivity look similar.
This is evidenced by the respective distributions of the $Q$ factor in such configurations (cf. Figs. \ref{f:f2}--\ref{f:f7}).
So we expect that in reality they will respond to evolving boundary conditions in a similar way by accumulating intense currents in the corona at approximately the same sites.
A detailed comparison of such processes in these types of
configurations will require the use of fully time-dependent MHD models
(e.g., \cite{Linker2010}).

In addition, it appears to us that focusing just on {\it photospheric} coronal hole connections may be somewhat misleading.
Even when coronal holes are only linked in the photosphere, we find that they connect robustly in the low corona along the separator field line.
This topological feature is extremely favorable for magnetic reconnection compared to isolated null points, because it can occur over the entire length of the separator rather than being confined to a small region around the nulls (a comprehensive analysis of this issue is recently given by  \citet{Parnell2010}).
Therefore, the most interesting processes are expected to develop at these separators and they will be in the focus of our future studies.

%
%
%
\section{IMPLICATIONS FOR THE CORONAL PHYSICS
	\label{s:impl}}

The considered example has several important implications for the coronal physics, particularly, for understanding the solar wind nature.
Let us start to discuss them by considering first the map of coronal holes calculated in MHD simulations of the corona during the time period of the total solar eclipse 2008 August 1 \citep{Rusin2010}.
This map is shown in Figure \ref{f:f9} (top panel), where the coronal holes are shaded  in dark blue and red at negative and positive polarities, respectively, and superimposed on the photospheric $B_r$ distribution that is used as input data for the MHD model.
One can clearly see from this panel that coronal holes of both polarity occupy multiply-connected domains with many apparently disconnected components at low latitudes.
Similarly to what we had in our simple example considered in the previous sections, the coronal holes are disconnected here by parasitic polarity regions of opposite sign compared to the holes.

Trying to verify our extended uniqueness conjecture in this eclipse case, we have found that the value
\begin{eqnarray}
  \slog Q \equiv \sign(B_r) \log\left[Q/2 + \left(Q^2/4-1\right)^{1/2}\right]
	\label{slogQ}
\end{eqnarray}
is very convenient for characterizing the photospheric coronal hole linkages.
Here the expression under the logarithm is an exact expression for the squashing factor ($\ge 1$) in terms of its asymptotic values $Q\ge 2$ \citep{Titov2002, Titov2007a}.
This value practically coincides at $Q\gg 2$  with $\log Q$ taken with the sign of $B_r$ at the boundary, so we call it {\it signed log $Q$} or simply $\slog Q$.
Applying the red-blue palette to the  $\slog Q$ distributions, we are able to simultaneously visualize (quasi-)separatrix footprints and the sign of the respective magnetic polarities.
The bottom panel in Figure \ref{f:f9} shows the photospheric  $\slog Q$ distribution for the eclipse case together with the coronal holes shaded in the same way as in the top panel.

Comparing these panels,  we see that there are several clusters of disconnected parts of the coronal holes that are indeed linked by high-$Q$ lines representing the footprints of separatrix or quasi-separatrix surfaces.
The linkage between the disconnected parts spreads out within a given cluster, reaching eventually the main polar hole of the same polarity.
The most obvious linkage of such a cluster to the hole can be seen above the equator at $\phi \leq 90\arcdeg$, where the cluster links shortly to the northern polar hole.
This cluster is delineated in Figure \ref{f:f9}, together with its associated linkages, by a thick transparent green line.

The other linkages are less obvious, but nevertheless recoverable from the presented $\slog Q$ distribution, except for one interesting case.
The latter refers to a single small hole located at $\phi \approx 280\arcdeg$ and $\theta \approx 20\arcdeg$ inside an isolated negative polarity, which is circled in Figure \ref{f:f9} by a transparent green line.
This hole is nested into a positive polarity region formed by a compact group of positive flux concentrations.
According to the nested conjecture by \citet{Antiochos2007}, such a hole itself must be nested into another coronal hole of the same sign.
This is indeed the case but only if we interpret the three neighboring regions of negative open field as one composite hole linked by high-$Q$ lines into a ``necklace" that encircles the isolated small hole and thus makes it nested, as required by the conjecture.
This ``necklace" links to the cluster of open negative field regions, whose high-$Q$ lines do not spread out farther than $\theta\approx 40\arcdeg$, because at these latitudes the indicated cluster becomes isolated from the northern polar hole by a very dense group of positive polarity regions.
One can check, however, that this cluster still links to the northern polar hole via high-$Q$ lines propagating in the corridor $150\arcdeg \leq \phi \leq 180\arcdeg$, as required by our extended conjecture.
Both this corridor and the cluster itself are delineated in Figure \ref{f:f9} by a transparent green line.

\begin{figure}
\epsscale{0.9}
\plotone{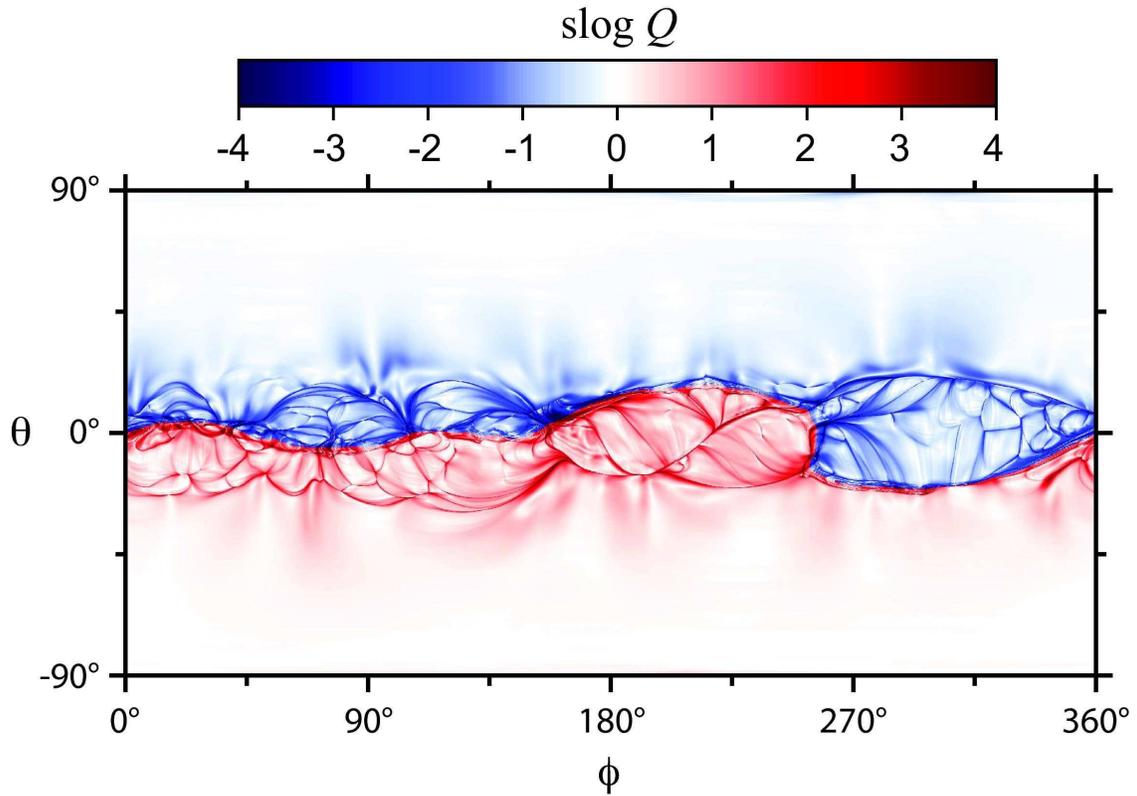}
\caption{The results of the solar wind model for the total solar eclipse 2008 August 1: the $\slog Q$ distribution (see equation (\ref{slogQ})) at the sphere $r=3 R_{\sun}$.  The high-$Q$ lines border here the regions of open magnetic flux that appear at the photosphere as disconnected or nearly disconnected ones. Blue and red colors correspond, respectively, to negative and positive magnetic fluxes.
	\label{f:f10} }
\end{figure}

Figure \ref{f:f10} presents the $\slog Q$ distribution at $r=3\Rs$ where the field structure becomes open all over the sphere, except for a narrow belt that follows the neutral line, widening no more than $10\arcdeg$.
The belt corresponds to the heliospheric current sheet, which consists of very stretched flux tubes in the radial direction that are still closed at $r=3\Rs$ flux tubes.
A comparison of this $\slog Q$ distribution with the calculated coronal holes shows that, as expected, a part of the high-$Q$ lines outlines the indicated belt.
Outside of the belt, this distribution reveals a very intricate network of high-$Q$ lines arching between different parts of the heliospheric current sheet; such a network was called {\it S-web} by  \citet{Antiochos2010}.
The example described in the previous sections strongly suggests that the S-web is formed by multiple (quasi-)separatrix curtains that fall down to the parasitic polarities and join the separatrix domes along (quasi-)separator field lines.
In this way, the S-web borders the fluxes of the individual components of coronal holes that are nearly disconnected at the photospheric level or linked by the footprints of the respective separatrix domes.
Our study indicates also that the connection between these components is restored at certain levels above the photosphere.
This prediction is indeed confirmed by our direct computations of the coronal holes at different radii in the present eclipse case \citep{Antiochos2010}. 
On the basis of our above example, we could also predict that the coronal hole junctions can occur approximately at the heights, where the null points, similar to the nulls N$_3$ and N$_2$ in our configuraiton, are located.
A detailed proof of this prediction goes far beyond the scope of the present paper, but our preliminary analysis of a few such junctions for the present eclipse case is in agreement with this prediction.

A more important conclusion that follows from our simple example is that a single parasitic polarity region can produce at the top of its separatrix dome several magnetic null points.
Depending on the number of nulls, we deal here either with the separator field line connecting two or three nulls or the quasi-separator if the null point is single but the parasitic polarity has an elongated shape.
These (quasi-)separators are manifested as high-$Q$ lines passing along the middle of the parasitic polarities intruded into coronal holes (see the bottom panel in Fig. \ref{f:f9}). 
As photospheric boundary conditions vary, the number of the nulls can change together with changing locations of the parasitic polarities, which automatically implies changing of the magnetic topology.
The latter can occur in MHD approach only via dynamic formation of strong current layers and subsequent reconnection of magnetic field in them.

So  we expect that, in reality, the existing continual variation of parasitic polarities triggers the reconnection between closed and open magnetic fields at the top of their separatrix domes where the (quasi-)separators are located.
Such an ongoing process then should be a persistent source of plasma outflows.
Since the (quasi-)separators represent also the bottom edge of (quasi-)separatrix curtains, we anticipate that the plasma outflows caused by reconnection can easily spread along open field lines high up into the corona.
Thus, this process appears to be promising for providing a substantial supply of the material for the solar wind and, particularly, for its slow component.
There are several supporting arguments in favor of this hypothesis, which are described in detail in our other papers  \citep{Linker2010, Antiochos2010}.
Here we would like to point out only one of them, namely, that the location of our separatrix curtains apparently fits well to the location of the so-called pseudo-streamers or plasma sheets \citep{Hundhausen1972, Neugebauer2002}.
These features are characterized by an enhanced plasma density and observed above unipolar magnetic regions separating the coronal holes of the same polarity, which is approximately at the place where our (quasi-)separatrix curtains are located.

Our analysis of coronal hole linkages is also of substantial interest for the physics of coronal mass ejections (CMEs).
Recently, \citet{Liu2006} and \citet{Liu2007b} have demonstrated that the fastest CMEs originate in the vicinity of the mentioned pseudo-streamers.
In the light of our present analysis, this conclusion looks very natural.
Indeed, imagine that the erupting flux appears, for example, as a result of magnetic field emergence from beneath the photosphere and partly inside a parasitic polarity region that splits the coronal hole into two parts.
Since the magnetic field is open above such a region, all closed magnetic flux overlying the newly emerging field is only due to this parasitic polarity.
In general, this flux does not seem to be large compared to the one that is usually found below the heliospheric current sheet, and hence its capacity to ``tether" the emerging field must be smaller.
Therefore, we expect that in this case the erupting flux can gain a faster propagation speed after its protrusion through the closed field of the parasitic polarity.
It also should be emphasized that, as in the breakout model \citep{Antiochos1999}, the opening of the closed field here can be alleviated by reconnection at the above (quasi-)separator field lines.

%
%

\section{SUMMARY \label{s:sum}}

In this paper, we have analyzed the variation of magnetic topology in response to the motion of a parasitic polarity region intruded into a bulge of the polar coronal hole.
First, we have constructed an exact analytical source surface model of the solar magnetic field by using the electrostatic method of images and Kelvin transform.
The model consists of three components: a global dipole-type field, a bipole active region, and an elongated negative flux spot, called the parasitic polarity.
The first two components produce the solar coronal field with asymmetric polar coronal holes, which are bulging toward the equator.

We start from the reference state, where a parasitic negative polarity is placed across a local bulge of the northern coronal hole that has a positive magnetic polarity, so that the bulge becomes completely disconnected from the main hole (Figs. \ref{f:f1}c and \ref{f:f2}).
This state, denoted as \Stoo, has three topological features in the neighborhood of the parasitic polarity: one bald patch (BP) and two magnetic null points N$_2$ and N$_1$ connected by a {\it separator} field line, which lies at the intersection of two fan surfaces.
The first fan surface emanates from the null N$_1$ and forms a separatrix {\it dome} that covers the parasitic polarity and completely isolates its flux from the surrounding field.
The field lines belonging to the dome are connected either to the BP or to the null N$_1$, except for a single spine line that connects to the null N$_2$ (Fig. \ref{f:f3}).
The second fan surface emanates from the null N$_2$ and forms a sort of separatrix {\it curtain} that borders the fluxes coming out from the disconnected bulge of the coronal hole and its main body.
The bordering begins at the height of the null N$_2$, where the disconnected parts of the coronal hole first come into contact with each other, and continues up to the source-surface, where all field lines become open.
Below the null N$_2$, the northern coronal hole remains formally {\it disconnected} but {\it linked} by the separatrix dome.

Withdrawing the parasitic polarity from the coronal hole in the westward direction, we gradually restore the connection inside the hole.
We have identified the basic topological states in this process, whose essence is in the consecutive transformation of the separatrix curtain into a quasi-separatrix one.
This transformation begins when a third null point N$^{\sun}_3$ appears at the BP (Fig. \ref{f:f4}) and starts rising up into the corona (Fig. \ref{f:f5}).
Initially, the spine field line emanating from this null is a closed loop, whose height, however, rises until its apex touches the source surface at the neutral line.
This happens at the state \Stoiis, where all the field lines entering the null $\hat{\rm N}_3$ connect to the source surface via that spine line.
Thus, exactly at this moment, the northern coronal hole becomes {\it singularly connected} at the photosphere through a line corridor, which coincides with the eastern part of the dome footprint.
Further rise of the null N$_3$ shifts its spine to the west, widens the corridor to a finite size, by turning it into a footprint of an HFT, whose upper QSL with open magnetic flux replaces a part of the separatrix curtain.
The null N$_3$ moves westward faster than the null N$_2$, so that they approach each other and eventually coalesce into a single degenerate null N$^*_2$ (Fig. \ref{f:f6}), which turns at the next instant into a magnetic minimum point.
The separator at this state turns into a {\it quasi-separator}, while the curtain becomes entirely quasi-separatrix (Fig. \ref{f:f7}).
This curtain then gradually shrinks to the west with further movement of the parasitic polarity in this direction (Fig. \ref{f:f8}).
The disconnection of this coronal hole occurs in the reverse order when the parasitic polarity is moving eastward from the reached position.
The most important moments in this process is the appearance of the indicated magnetic minimum above the parasitic polarity and its subsequent bifurcation into a pair of null points.

In a highly conducting coronal plasma, the described transformation of the configuration must occur via the formation of a current layer and reconnection over the entire (quasi-)separator field line (see \citet{Parnell2010}) rather than being confined to small regions around the nulls.
On the basis of this consideration, we argue that the respective reconnection outflows along and nearby separatrix curtains may serve as a substantial source of the slow solar wind.
The configurations with the parasitic polarities splitting coronal holes into two parts must also be favorable for the eruption and propagation of unstable magnetic structures in the solar corona.


\acknowledgments

We thank the referee for a careful review of our paper, which has helped us to significantly improve the
presentation of our results.
The contribution of V.S.T., Z.M., J.R.L., and R.L. was supported by
NASA's Heliophysics Theory, Living With a Star, and SR\&T programs,
and the Center for Integrated Space Weather Modeling (an NSF Science
and Technology Center). The contribution by S.K.A. was supported by
the NASA HTP, TR\&T, and SR\&T programs.


\appendix

\section{CALCULATION OF MAGNETIC NULL POINTS
	\label{s:apx}}
	
The null points presented in Table \ref{t:1} have been calculated with an accuracy to nine significant digits by using computer-algebraic system Maple.
First, for each topologically stable state we have determined graphically the number of null points and their approximate coordinates.
This was done simply by finding the points where all three iso-surfaces $B_r=0$, $B_\theta=0$, and $B_\phi=0$ intersect each other.
The graphical method allowed us to determine at least two significant digits of the null-point coordinates.
Then, using these coordinates as the initial ones, we have reached the indicated nine significant digits iteratively.

Since our exact analytical expression of $\bm B$ is rather complicated, its direct use in the calculation of nulls as roots of the equation ${\bm B}=0$ is not efficient.
Therefore, at each iteration we approximated first this expression by its second-order Taylor expansion about the root that has been obtained in the previous iteration.
Then, using this expansion, we calculated its root by a Maple procedure based on the Newton-Raphson algorithm.
Repeating these operations several times, we reached the indicated accuracy, which has been checked by direct substitution of the found nulls into our exact expression of $\bm B$.

\begin{figure}
\epsscale{0.9}
\plotone{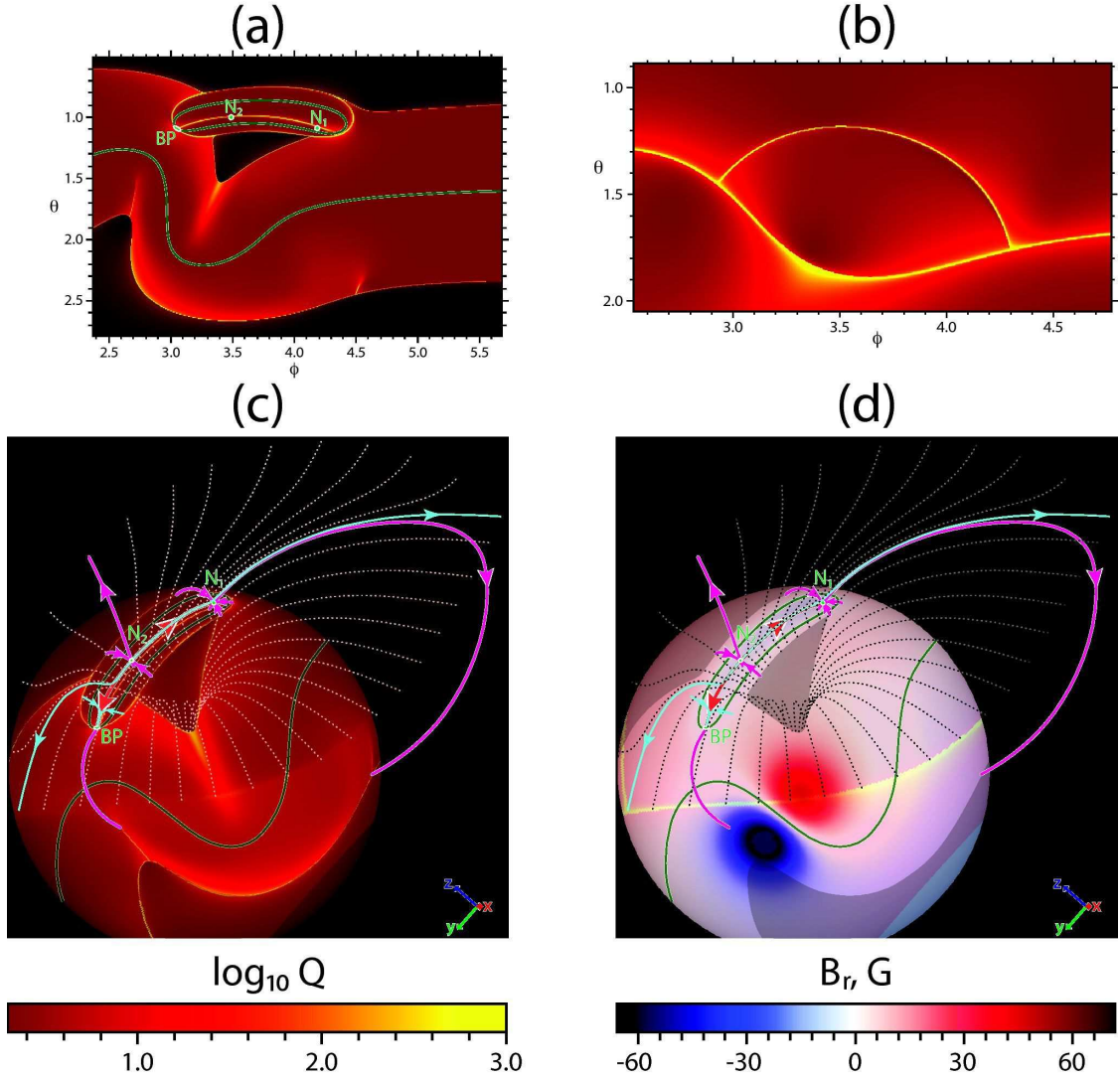}
\caption{ 
\setlength{\baselineskip}{16pt}   
The reference topological state \Stoo with a bald patch (BP) and two magnetic nulls $\rm N_2$ and $\rm N_1$ located above the parasitic polarity: the $Q$ distributions in the $(\theta,\phi)$-plane at the photosphere (a) and the source surface (b), respectively, and the corresponding topological skeleton of the magnetic field shown in panels (c) and (d) together with the respective photospheric $Q$ and $B_r$ distributions.
The semi-transparent grey-shaded areas indicate at the photosphere the coronal holes.
The thick green lines represent the photospheric polarity inversion lines.
The scarlet and magenta thick lines depict, respectively, the separator and other separatrix lines that emanate from the nulls exactly along their eigenvectors.
The solid cyan lines represent either the BP separatrix lines (short lines) or the fan separatrix field lines (associated with the null $\rm N_2$ in this case), while the semi-transparent dotted field lines show the boundary of the disconnected coronal hole.
The $Q$ distribution (b) is mapped in panels (c) and (d) on the source surface $r=2.5 R_\sun$ by using semi-transparent colors.
The vector triad in the lower right-hand corner of panels (c) and (d) indicates the angle orientation of the Cartesian coordinate system that is rigidly bound to the Sun center with the $z$-axis passing through the poles.
	\label{f:f2} }
\end{figure}

\begin{figure}
\epsscale{0.95}
\plotone{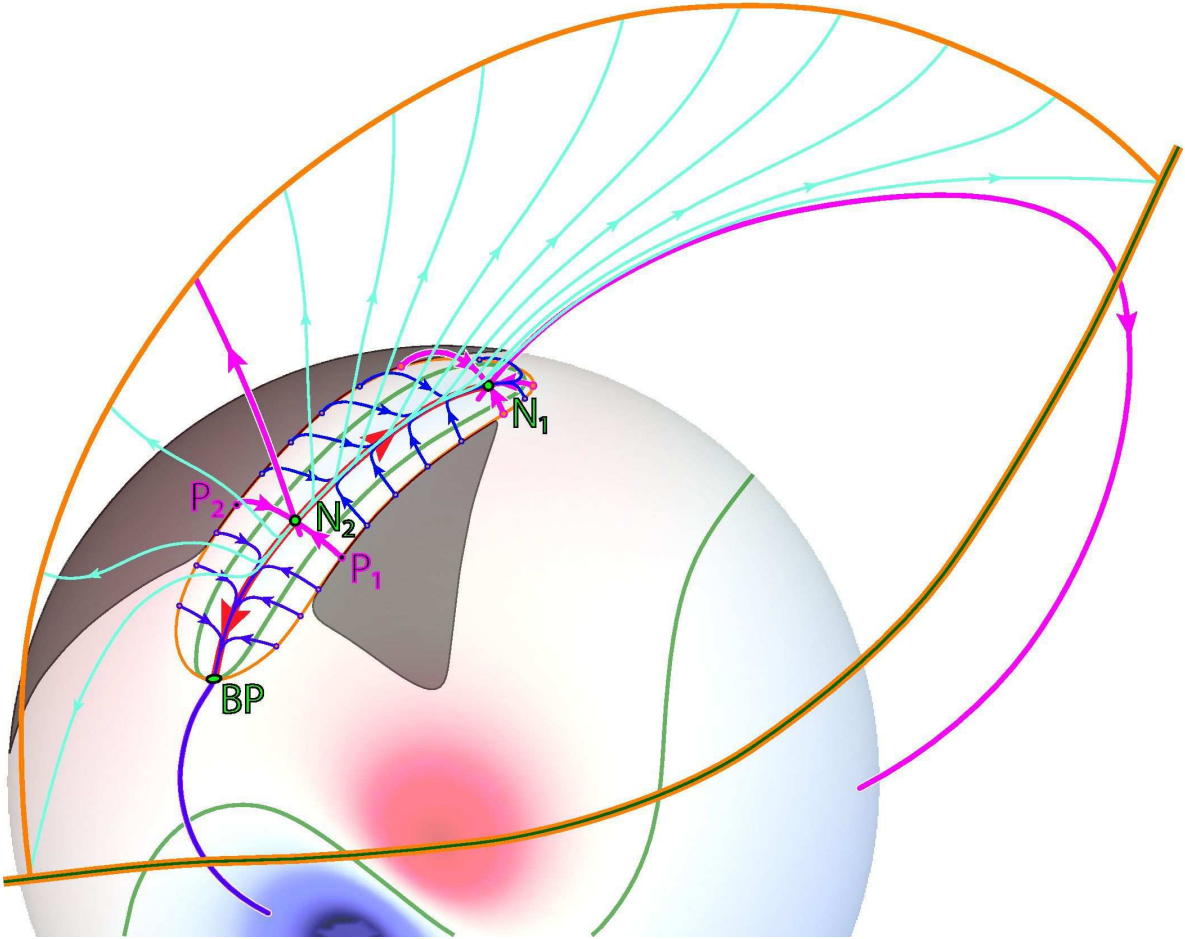}
\caption{The structure of the magnetic field lines at the separatrix curtain and dome in the topological state \Stoo (see Fig. \ref{f:f2}).
All the field lines of the separatrix curtain (cyan lines) emanate from the null point N$_2$, while the field lines of the separatrix dome connect either to the BP (purple lines) or to the null N$_1$ (blue lines).
Such field lines converge and propagate very closely to the separator (thick scarlet line), which threads two nulls and touches the BP.
The ``purple" and ``blue" subsets of field lines are separated by the spine line (thick  magenta line) of the null N$_2$.
The footpoints P$_1$ and P$_2$ of this spine line are the only points at which the dome footprint (thin orange line) touches the disconnected parts of the coronal hole.
At the source surface, the thick orange line represents the footprint of the separatrix curtain, while the green line with an orange ``aura" depicts the neutral line, i.e. the base of the heliospheric current sheet.
	\label{f:f3} }
\end{figure}

\begin{figure}
\epsscale{0.9}
\plotone{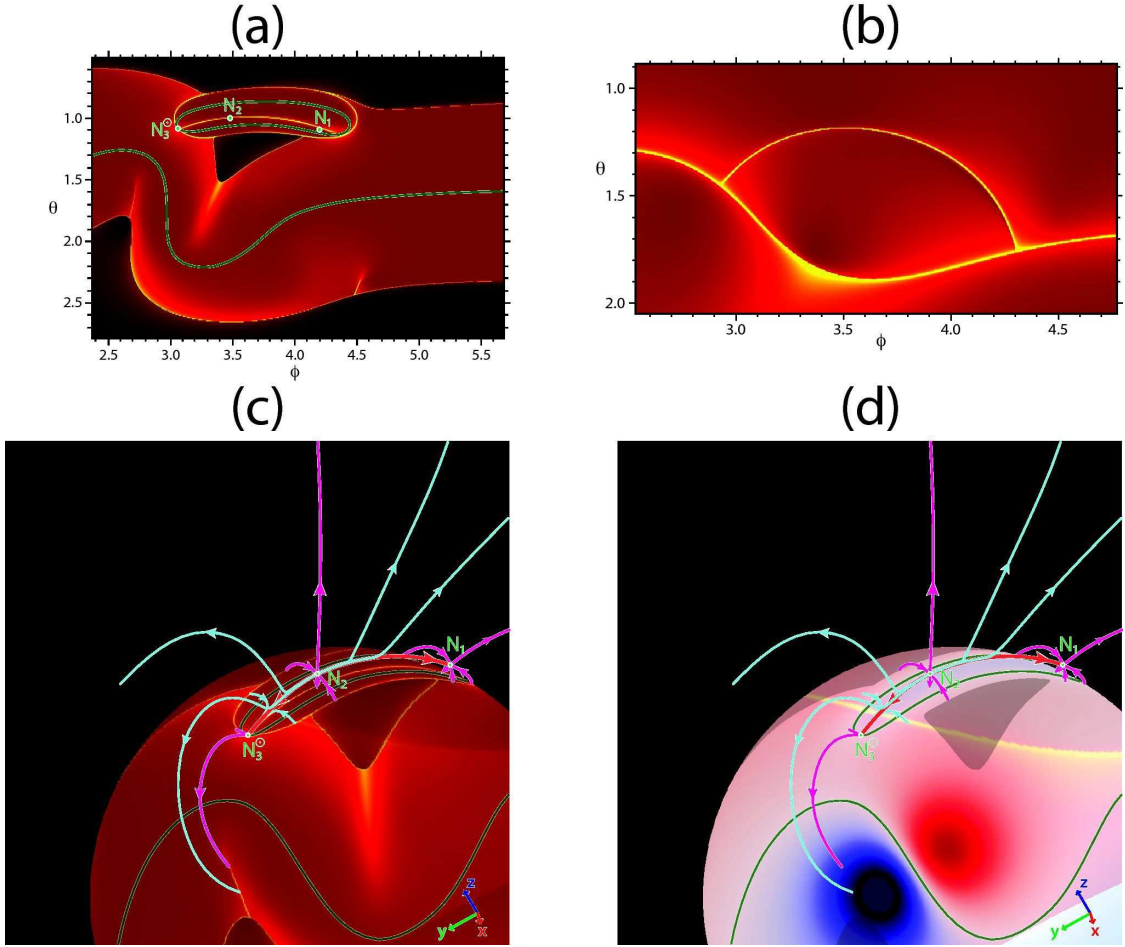}
\caption{The topological state \Stoi with three magnetic nulls, one of which ($\rm N^{\sun}_3$) is located at the photosphere, while the other two ($\rm N_2$ and $\rm N_1$) are above the parasitic polarity.
All three null points are connected by a separator field line.
The panels and color coding of the lines and surfaces are the same as in Figure \ref{f:f2}, where the used color bars are shown too.
Other field lines are not shown here, since they are similar to those depicted in Figure  \ref{f:f2}.
	\label{f:f4} }
\end{figure}

\begin{figure}
\epsscale{0.9}
\plotone{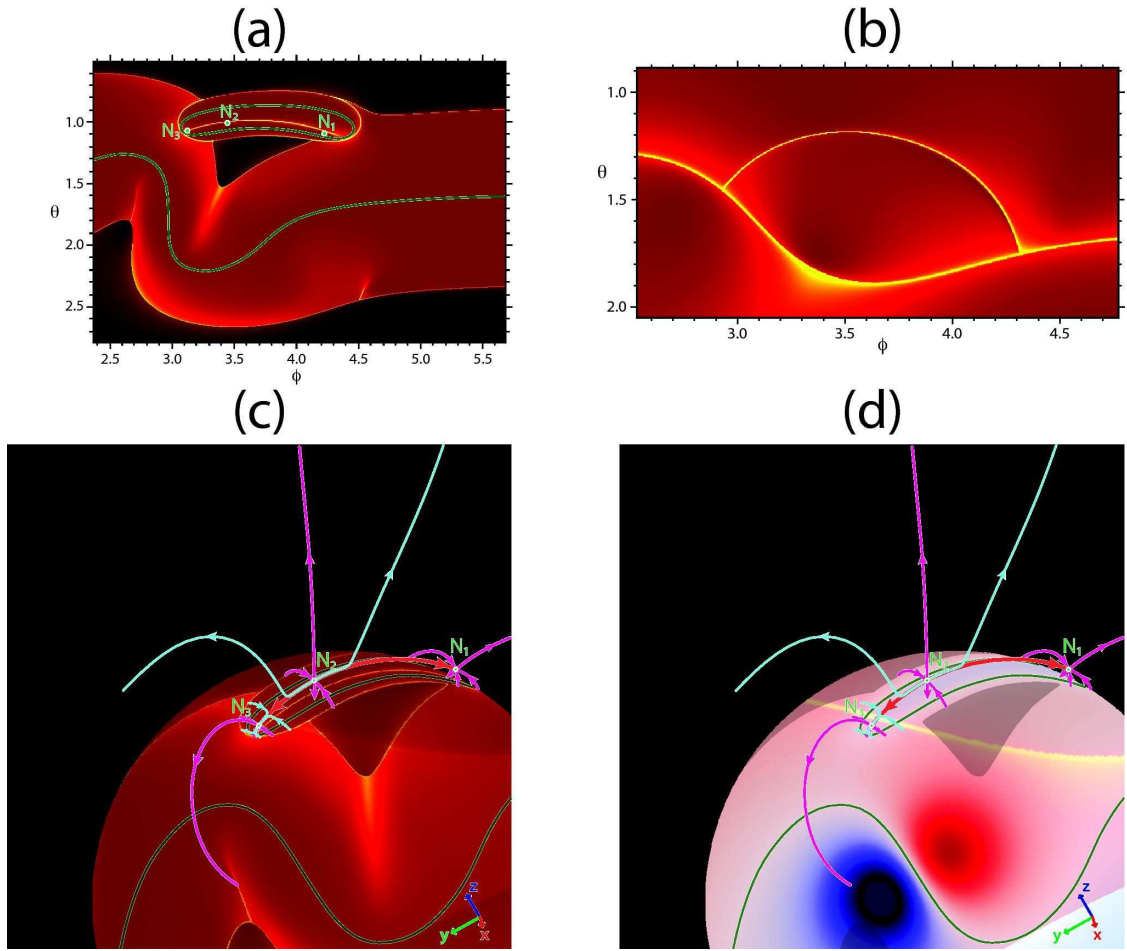}
\caption{The topological state \Stoii with three magnetic nulls $\rm N_3$, $\rm N_2$, and $\rm N_1$, all located above the parasitic polarity.
All three null points are connected by a separator field line.
The panels and color coding of the lines and surfaces are the same as in Figure \ref{f:f2}, where the used color bars are shown too.
Other field lines are not shown here, since they are similar to those depicted in Figure  \ref{f:f2}.
	\label{f:f5} }
\end{figure}

\begin{figure}
\epsscale{0.9}
\plotone{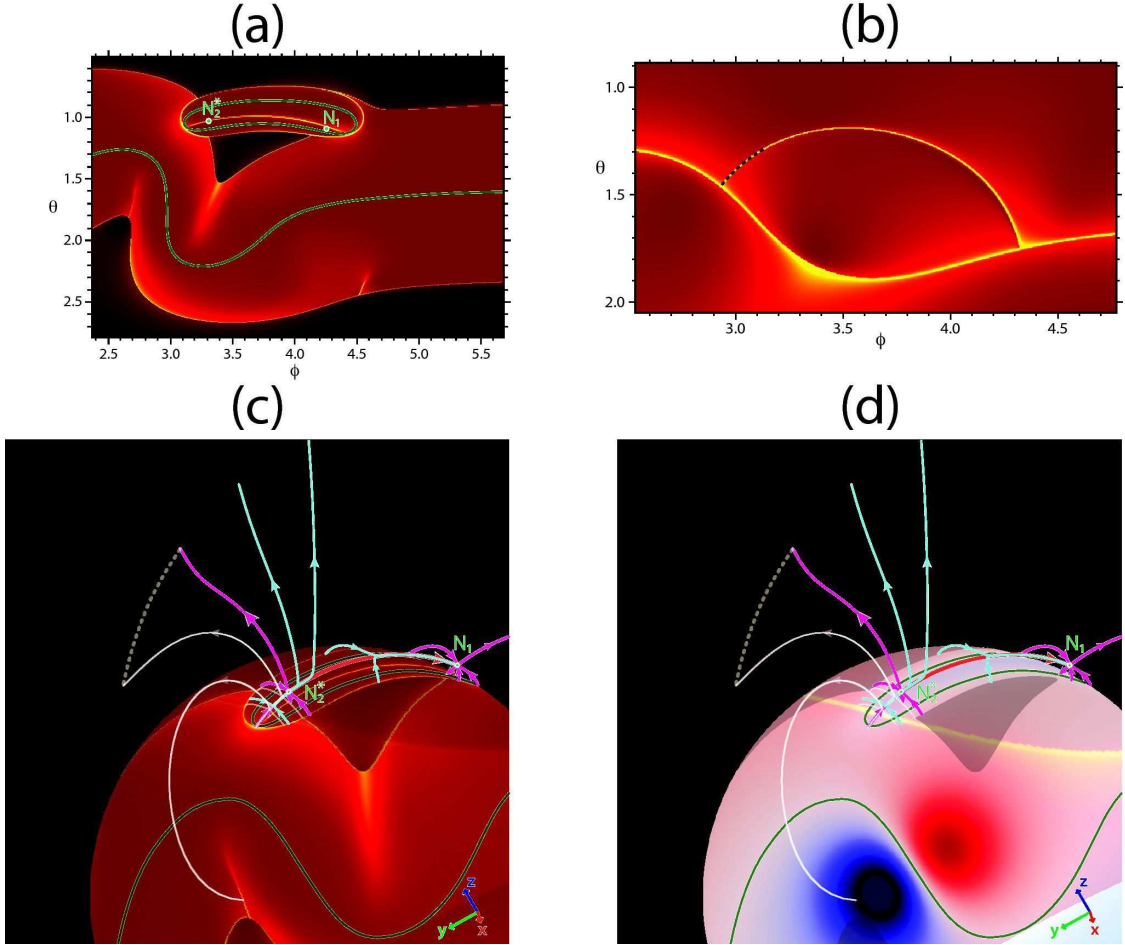}
\caption{The topological state \Stoiii with a hyperbolic flux tube (HFT) and two magnetic nulls N$^*_2$ and N$_1$, all located above the parasitic polarity on a separator field line.
The null N$^*_2$ is degenerate in the sense that one of its eigenvalues identically vanishes.
The panels and color coding of the lines and surfaces are the same as in Figure \ref{f:f2}, except of the new type of field lines that are colored here in semi-transparent white: they belong to an HFT, whose photospheric footprint forms an extremely narrow corridor connecting the initially disconnected parts of the northern coronal hole.
The source-surface footprint of the HFT is shown by dashed lines: black one on panel (b) and white one on panels (c) and (d).
Other field lines are not shown here, since they are similar to those depicted in Figure  \ref{f:f2} (see this figure for color bars too).
	\label{f:f6} }
\end{figure}

\begin{figure}
\epsscale{0.9}
\plotone{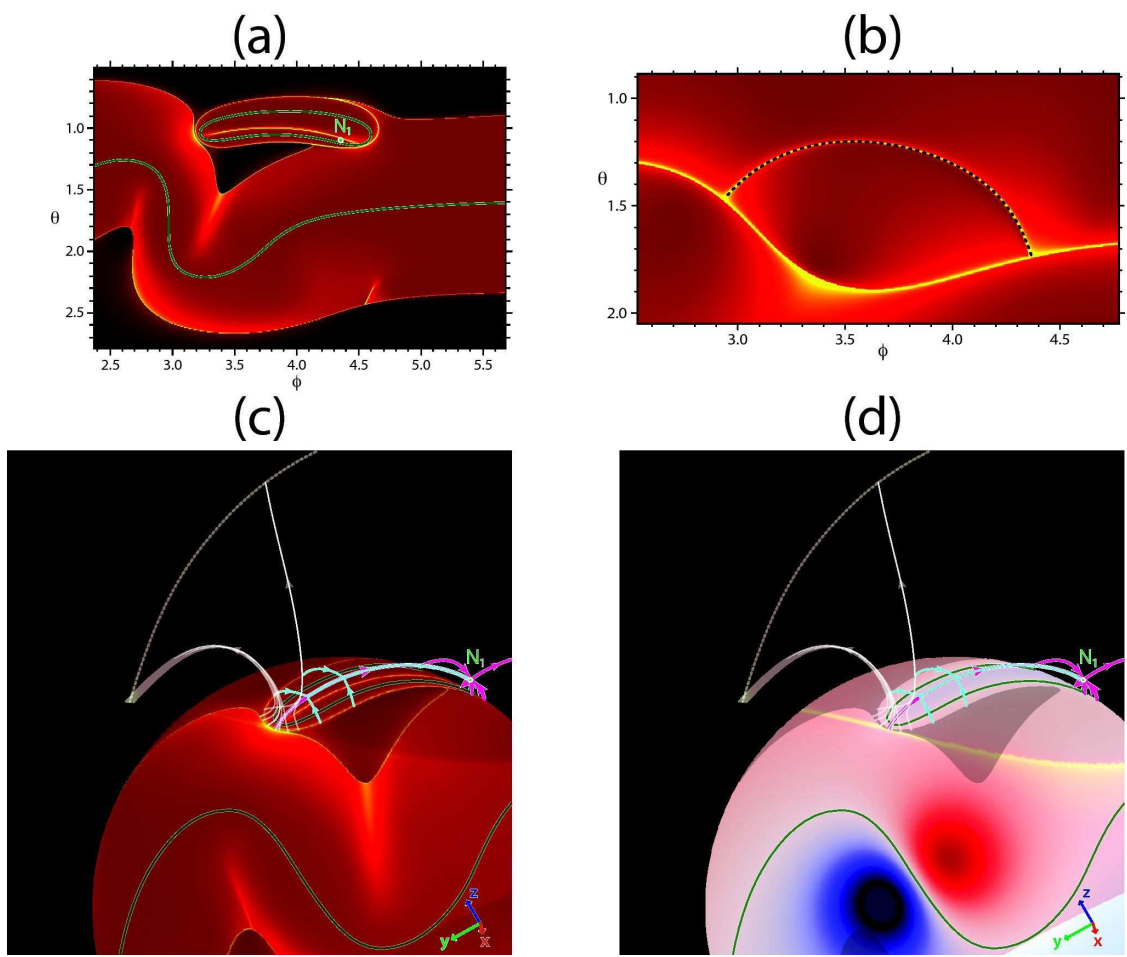}
\caption{The topological state \Stoiv with an HFT and a magnetic null N$_1$ located above the parasitic polarity.
The panels and color coding of the lines and surfaces are the same as in Figures \ref{f:f2} and  \ref{f:f6}.
Other field lines are not shown here, since they are similar to those depicted in Figure  \ref{f:f2} (see this figure for color bars too).
In this state, a part of the northern coronal hole is nearly disconnected, so that only a narrow corridor link it at the photospheric level to the major part of the coronal hole.
This corridor is the photospheric footprint of the indicated HFT; its conjugate source-surface footprint is located along the high-$Q$ arc, traced by dashed black and white lines in panels (b) and (c) and (d), respectively.
	\label{f:f7} }
\end{figure}

\begin{figure}
\epsscale{0.9}
\plotone{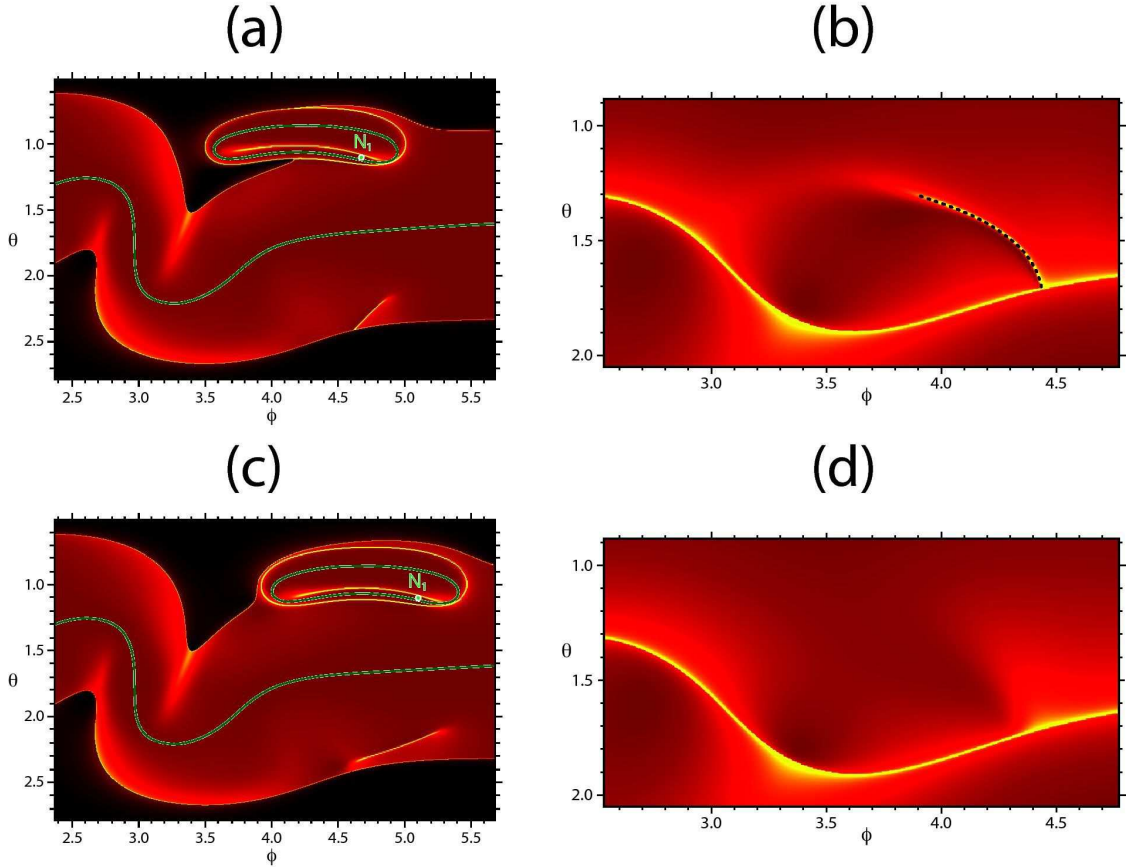}
\caption{Topological states \Sti (panels (a) and (b)) and \Stii (panels (c) and (d)) with and without an HFT, respectively.
The parasitic polarity is covered in both these states by a separatrix dome associated with the null point  N$_1$ that locates above the western part of the parasitic polarity.
In the former state, the parasitic polarity is on the half way to disconnect the bulge of the northern coronal hole (a); the eastern side of the parasitic polarity is skirted by an HFT.
Its source-surface footprint appears as a high-$Q$ arc, whose western tip is anchored at the neutral line;  this arc is traced by a dashed black line in panel (b).
	\label{f:f8} }
\end{figure}

\begin{figure}
\epsscale{0.9}
\plotone{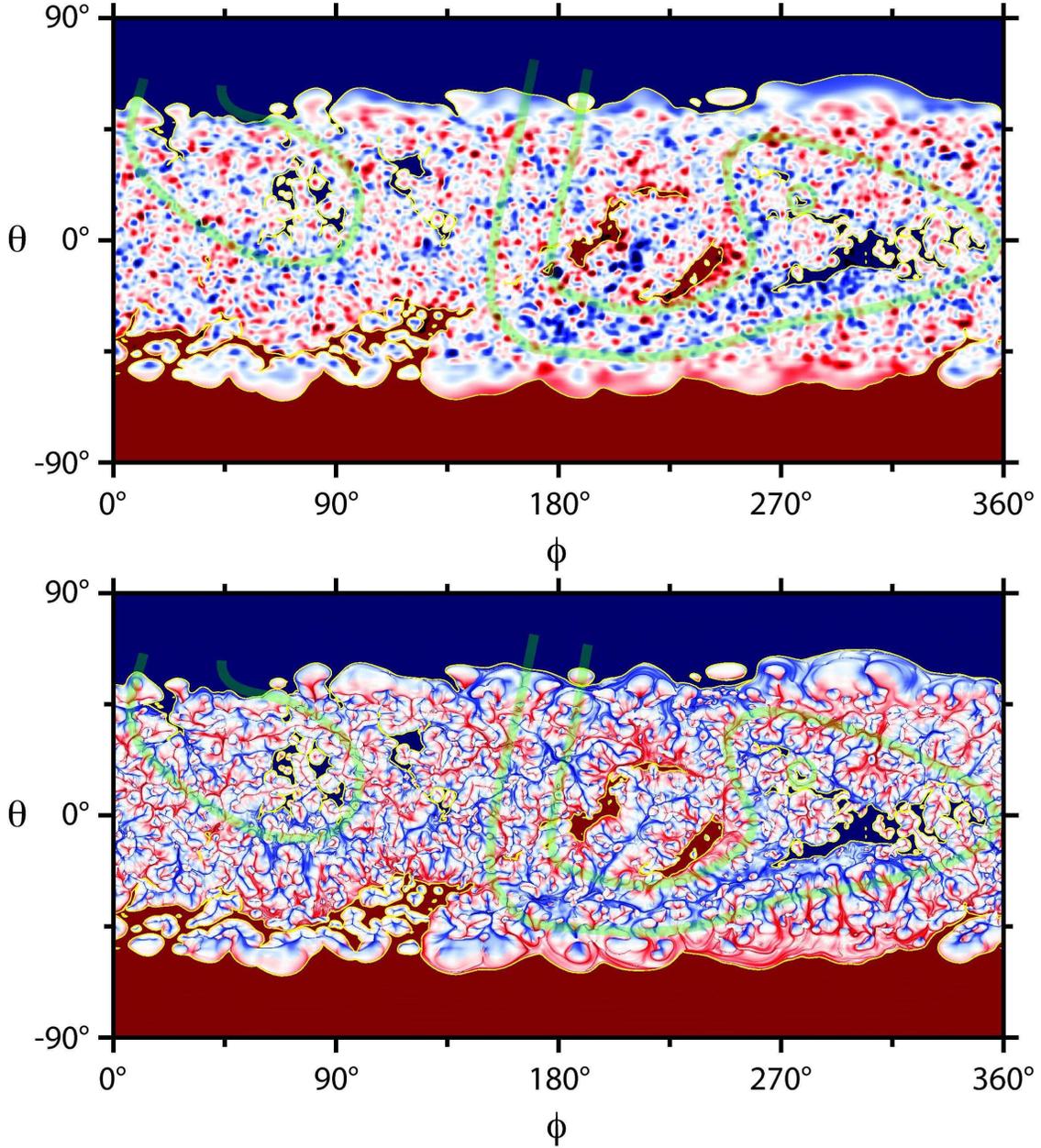}
\caption{The results of the solar wind model for the total solar eclipse 2008 August 1: The photospheric $B_r$ (top panel) and $\slog Q$ (bottom panel, see equation (\ref{slogQ})) distributions; the coronal holes are shaded in both panels in dark red and blue.
Color scale bars for $B_r$  and $\slog Q$ are the same as in Figure \ref{f:f2} and \ref{f:f10}, respectively, except that $B_r$ here is saturated approximately at $\pm20\:$G. 
The transparent green lines delineate the clusters of disconnected coronal holes discussed in Section \ref{s:impl} and the adjacent regions through which these clusters are linked by high-$Q$ lines to the respective main hole. 
	\label{f:f9} }
\end{figure}

The obtained null points have been used for the analysis of the local field structure, which included, in particular, the determination of eigenvalues and eigenvectors of the matrix $\left[ \bm \nabla \bm B \right]$ evaluated at these points.
This analysis provided us with the required information for plotting the topological skeleton of our configuration at different states, shown in Figures \ref{f:f2}--\ref{f:f8}.

To determine topologically unstable states, we have generalized the described procedure by regarding the parameter $\phi_a$ as a fourth unknown in addition to the previous three ones and properly extending the system $\bm B =0$ with an extra equation.
For the state \Stoi, the extra equation is $r=R_{\sun}$, which simply requires that the null N$^{\sun}_3$ must be located exactly at the photosphere.
For the state \Stoiii, the extra equation is  $\det\left[ \bm \nabla \bm B \right]=0$, which is the necessary and sufficient condition of vanishing of one of the eigenvalues of the matrix $\left[ \bm \nabla \bm B \right]$ at the null point N$^*_2$.
The respective values of $\phi_a$ and the null-point coordinates were determined for these states  with the same accuracy as for the others.



\begin{thebibliography}{33}
\expandafter\ifx\csname natexlab\endcsname\relax\def\natexlab#1{#1}\fi

\bibitem[{{Altschuler} \& {Newkirk}(1969)}]{Altschuler1969}
{Altschuler}, M.~D., \& {Newkirk}, G. 1969, \solphys, 9, 131

\bibitem[{{Antiochos} {et~al.}(2007){Antiochos}, {DeVore}, {Karpen}, \&
  {Miki{\'c}}}]{Antiochos2007}
{Antiochos}, S.~K., {DeVore}, C.~R., {Karpen}, J.~T., \& {Miki{\'c}}, Z. 2007,
  \apj, 671, 936

\bibitem[{{Antiochos} {et~al.}(1999){Antiochos}, {DeVore}, \&
  {Klimchuk}}]{Antiochos1999}
{Antiochos}, S.~K., {DeVore}, C.~R., \& {Klimchuk}, J.~A. 1999, \apj, 510, 485

\bibitem[{{Antiochos} {et~al.}(2010){Antiochos}, {Miki{\'c}}, {Lionello},
  {Titov}, \& {Linker}}]{Antiochos2010}
{Antiochos}, S.~K., {Miki{\'c}}, Z., {Lionello}, R., {Titov}, V.~S., \&
  {Linker}, J.~A. 2010, \apj, submitted

\bibitem[{{Axler} {et~al.}(2001){Axler}, {Bourdon}, \& {Ramey}}]{Axler2001}
{Axler}, S.~J., {Bourdon}, P., \& {Ramey}, W. 2001, Graduate texts in
  mathematics, Vol. 137, {Harmonic function theory} (Berlin: Springer, 259 p.)

\bibitem[{Baker {et~al.}(2009)Baker, van Driel-Gesztelyi, Mandrini, Demoulin, ,
  \& Murray}]{Baker2009}
Baker, D., van Driel-Gesztelyi, L., Mandrini, C.~H., Demoulin, P., , \& Murray,
  M.~J. 2009, \apj, 705, 926

\bibitem[{{Baum} \& {Bratenahl}(1980)}]{Baum1980}
{Baum}, P.~J., \& {Bratenahl}, A. 1980, \solphys, 67, 245

\bibitem[{{Bungey} {et~al.}(1996){Bungey}, {Titov}, \& {Priest}}]{Bungey1996}
{Bungey}, T.~N., {Titov}, V.~S., \& {Priest}, E.~R. 1996, \aap, 308, 233

\bibitem[{{D{\'e}moulin} {et~al.}(1996){D{\'e}moulin}, {Henoux}, {Priest}, \&
  {Mandrini}}]{Demoulin1996}
{D{\'e}moulin}, P., {Henoux}, J.~C., {Priest}, E.~R., \& {Mandrini}, C.~H.
  1996, \aap, 308, 643

\bibitem[{Harra {et~al.}(2008)Harra, Sakao, Mandrini, Hara, Imada, Young, van
  Driel-Gesztelyi, \& Baker}]{Harra2008}
Harra, L.~K., Sakao, T., Mandrini, C.~H., Hara, H., Imada, S., Young, P.~R.,
  van Driel-Gesztelyi, L., \& Baker, D. 2008, \apjl, 676, L147

\bibitem[{{Hundhausen, A.~J.}(1972)}]{Hundhausen1972}
{Hundhausen, A.~J.}, ed. 1972, Physics and Chemistry in Space, Vol.~5, {Coronal
  Expansion and Solar Wind} (Berlin: Springer, 238 p.)

\bibitem[{{Jackson}(1962)}]{Jackson1962}
{Jackson}, J.~D. 1962, {Classical Electrodynamics} (New York: Wiley, 808 p.)

\bibitem[{Kahler \& Hudson(2002)}]{Kahler2002}
Kahler, S.~W., \& Hudson, H.~S. 2002, \apj, 574, 467

\bibitem[{Ko {et~al.}(2006)Ko, Raymond, Zurbuchen, Riley, Raines, \&
  Strachan}]{Ko2006}
Ko, Y.-K., Raymond, J.~C., Zurbuchen, T.~H., Riley, P., Raines, J.~M., \&
  Strachan, L. 2006, \apj, 646, 1275

\bibitem[{{Landau} \& {Lifshitz}(1960)}]{Landau1960ecm}
{Landau}, L.~D., \& {Lifshitz}, E.~M. 1960, {Electrodynamics of continuous
  media} (Pergamon Press ; Addison-Wesley, Oxford : Reading, Mass. :), 417 p.

\bibitem[{{Lau} \& {Finn}(1990)}]{Lau1990}
{Lau}, Y.-T., \& {Finn}, J.~M. 1990, \apj, 350, 672

\bibitem[{{Linker} {et~al.}(2010){Linker}, {Lionello}, {Miki{\'c}}, {Titov}, \&
  {Antiochos}}]{Linker2010}
{Linker}, J.~A., {Lionello}, R., {Miki{\'c}}, Z., {Titov}, V.~S., \&
  {Antiochos}, S.~K. 2010, \apj, submitted

\bibitem[{{Liu}(2007)}]{Liu2007b}
{Liu}, Y. 2007, \apjl, 654, L171

\bibitem[{{Liu} \& {Hayashi}(2006)}]{Liu2006}
{Liu}, Y., \& {Hayashi}, K. 2006, \apj, 640, 1135

\bibitem[{{Longcope}(2001)}]{Longcope2001}
{Longcope}, D.~W. 2001, Phys.~Plasmas, 8, 5277

\bibitem[{{Neugebauer} {et~al.}(2002){Neugebauer}, {Liewer}, {Smith}, {Skoug},
  \& {Zurbuchen}}]{Neugebauer2002}
{Neugebauer}, M., {Liewer}, P.~C., {Smith}, E.~J., {Skoug}, R.~M., \&
  {Zurbuchen}, T.~H. 2002, \jgr, 107, 1488

\bibitem[{Parnell {et~al.}(2010)Parnell, Haynes, \& Galsgaard}]{Parnell2010}
Parnell, C.~E., Haynes, A.~L., \& Galsgaard, K. 2010, J. Geophys. Res., 115

\bibitem[{{Priest} \& {D{\'e}moulin}(1995)}]{Priest1995}
{Priest}, E.~R., \& {D{\'e}moulin}, P. 1995, \jgr, 100, 23443

\bibitem[{{Priest} {et~al.}(1996){Priest}, {Lonie}, \& {Titov}}]{Priest1996b}
{Priest}, E.~R., {Lonie}, D.~P., \& {Titov}, V.~S. 1996, J. Plasma Phys., 56,
  507

\bibitem[{{Priest} \& {Titov}(1996)}]{Priest1996}
{Priest}, E.~R., \& {Titov}, V.~S. 1996, Philos. Trans. R. Soc. London A, 354,
  2951

\bibitem[{{Ru{\v s}in} {et~al.}(2010){Ru{\v s}in}, {Druckm{\"u}ller}, {Aniol},
  {Minarovjech}, {Saniga}, {Miki{\'c}}, {Linker}, {Lionello}, {Riley}, \&
  {Titov}}]{Rusin2010}
{Ru{\v s}in}, V. {et~al.} 2010, \aap, 513, A45

\bibitem[{{Schatten} {et~al.}(1969){Schatten}, {Wilcox}, \&
  {Ness}}]{Schatten1969}
{Schatten}, K.~H., {Wilcox}, J.~M., \& {Ness}, N.~F. 1969, \solphys, 6, 442

\bibitem[{{Seehafer}(1986)}]{Seehafer1986a}
{Seehafer}, N. 1986, \solphys, 105, 223

\bibitem[{{Titov}(2007)}]{Titov2007a}
{Titov}, V.~S. 2007, \apj, 660, 863

\bibitem[{{Titov} {et~al.}(1999){Titov}, {D{\'e}moulin}, \&
  {Hornig}}]{Titov1999a}
{Titov}, V.~S., {D{\'e}moulin}, P., \& {Hornig}, G. 1999, in ESA SP-448:
  Magnetic Fields and Solar Processes, ed. A.~{Wilson} \& {et al.}, 715--722

\bibitem[{Titov {et~al.}(2009)Titov, Forbes, Priest, Mikic, , \&
  Linker}]{Titov2009}
Titov, V.~S., Forbes, T.~G., Priest, E.~R., Mikic, Z., , \& Linker, J.~A. 2009,
  \apj, 693, 1029

\bibitem[{{Titov} {et~al.}(2002){Titov}, {Hornig}, \&
  {D{\'e}moulin}}]{Titov2002}
{Titov}, V.~S., {Hornig}, G., \& {D{\'e}moulin}, P. 2002, \jgr, 107, 1164

\bibitem[{{Titov} {et~al.}(1993){Titov}, {Priest}, \&
  {D\'{e}moulin}}]{Titov1993}
{Titov}, V.~S., {Priest}, E.~R., \& {D\'{e}moulin}, P. 1993, \aap, 276, 564

\end{thebibliography}



\clearpage



%
%
%

\clearpage

\begin{deluxetable}{rllcccc}
\tablecolumns{7} 
\tablewidth{0in} 
\tablecaption{Basic topological states of the modeled configuration\tablenotemark{a}. \label{t:1}} 
\tablehead{ 
\colhead{State\tablenotemark{b}}    
&  \multicolumn{1}{c}{Figure} 
& \multicolumn{1}{c}{$\phi_{a}$}
& \multicolumn{1}{c}{\vspace{4pt}\rNn{r}{\theta}{\phi}{\rm N_{3}}}
& \multicolumn{1}{c}{\rNn{r}{\theta}{\phi}{\rm N_{2}}}
& \multicolumn{1}{c}{\rNn{r}{\theta}{\phi}{\rm N_{1}}}
& \multicolumn{1}{c}{\parbox{5.8em}{\small \baselineskip=1em   
                                 \hbox{} connection\tablenotemark{c}\\/ stability\tablenotemark{d}}}
 }
\startdata 
\Stoo	& \ref{f:f2}, \ref{f:f3}	& 3.72	 & ---						 	& \rN{1.0595}{1.0154}{3.4995}	& \rN{1.0446}{1.1050}{4.1834}   & L/S \\  
\Stoi		& \ref{f:f4}	 	  		& 3.7339	 & \rN{1.0}{1.1002}{3.0765}	 	& \rN{1.0595}{1.0176}{3.4877}	& \rN{1.0444}{1.1057}{4.1969}   & L/U \\
\Stoii		& \ref{f:f5}		  		& 3.76	 & \rN{1.0191}{1.0841}{3.1391}	& \rN{1.0592}{1.0232}{3.4562}	& \rN{1.0439}{1.1070}{4.2221}   & L/S \\
\Stoiis	& ---			  		& ?		 & ?							& ?						& ?						   & $\rm{\hat C}$/U \\
\Stoiii   	& \ref{f:f6}		  		& 3.7923	 & ---							& \rN{1.0521}{1.0460}{3.3225}	& \rN{1.0434}{1.1084}{4.2531}   & C/U \\  
\Stoiv	& \ref{f:f7}		  		& 3.892	 & ---							& ---						& \rN{1.0422}{1.1118}{4.3475}   & C/S \\ 
\Sti 		& \ref{f:f8}	(a, b)  		& 4.23 	 & ---							& ---						& \rN{1.0402}{1.1171}{4.6624}   & C/S \\ 
\Stii		& \ref{f:f8}	(c, d)  		& 4.68	 & ---							& ---						& \rN{1.0400}{1.1180}{5.0830}   & C/S \\
\enddata

\tablenotetext{a}{Parameters: $\Rss=2.5$, $m = 2.4224$, $|\pm q| = 2.4721$, $\mu = 0.10019$,
$d_{q} = 0.23980$, $d_{-q} = 0.21800$, $d_{a} = 0.2$, $l = 0.5$,  $\theta_{q}=\theta_{-q} = 1.765$, $\phi_{q} = 3.086$, $\phi_{-q} = 2.84$, $\theta_{a} =0.96$.  }
\tablenotetext{b}{Structural features: bald patch (BP), null point (N), and hyperbolic flux tube (HFT).}
\tablenotetext{c}{Coronal hole connection: connected (C), singularly connected ($\hat{\rm C}$), or only linked (L).}
\tablenotetext{d}{Topologically stability of the configuration: stable (S) or unstable (U).}
 
\end{deluxetable} 
%


\end{document}